# Survey on Data-Centric based Routing Protocols for Wireless Sensor Networks

Khalid S. Al Rasbi[1], Hothefa Shaker[1], Zeyad T. Sharef[2]

[1] Modern College of Business and Science, MCBS, AL-Khuwair 133, Sultanate of Oman
[2] College of Engineering, Ahlia University, Manama, Kingdom of Bahrain

**Abstract—** *The great concern for energy that grew with the technological advances in the field of networks and especially in sensor network has triggered various approaches and protocols that relate to sensor networks. In this context, the routing protocols were of great interest. The aim of the present paper is to discuss routing protocols for sensor networks. This paper will focus mainly on the discussion of the data-centric approach (COUGAR, rumor, SPIN, flooding and Gossiping), while shedding light on the other approaches occasionally. The functions of the nodes will be discussed as well. The methodology selected for this paper is based on a close description and discussion of the protocol. As a conclusion, open research questions and limitations are proposed to the reader at the end of this paper.*

## I. INTRODUCTION

In the recent years there have been great technological breakthroughs that relate to micro systems. This emphasis on the micro systems gave birth to new technologies and especially the invention of the micro sensors that have great communication capacities and abilities. They can process and communicate data efficiently. The sensors have certain circuits that are in direct contact with the environment that exists around the sensor itself. These circuits are intelligent enough to measure the environmental conditions and then change them into signals that are electric (Alipio and Tiglao 2017).

When the sensor receives the data collected by the circuits, it uses its radio transmitter to transfer that data to the command center, also known as sink. This process is done either in a direct transmission way or via data concentration center (DCC), also known as gateway. The sensor therefore has been considered as a powerful tool in data collection. Research has been intensive to see the possibilities of matching many sensors to work together within a unified network. The efforts made by various researchers have resulted in great benefits of using sensors nodes through a unified network. These benefits could be traced in various fields such as the military, the weather, surveillance, and security (Norouzi and Zaim, 2012) (Mondal and Sarddar, 2014).

The problem with using the sensor nodes lies in the fact that these sensors do face problems related to energy and bandwidth mainly. Research has therefore focused attention on the power awareness at system-level. The aim is to allow data relaying from the sensor nodes to the sink in a very efficient way. Therefore, this will maximize the network's lifetime. This is done through the use of the radio communication hardware, system partitioning, dynamic voltage scaling, and various other tools.

One of the problems that researchers face at this level is that to sensor networks it is not possible to apply classical IP-based protocols. Moreover, the sensor network applications need data flow coming from various sources to the sink. On the other hand, the same data could be communicated to the sink by various sensors. This uses energy and bandwidth whereas the results are rather redundant.

The solutions to these problems and limitations have manifested in the creation of new algorithms. The routing protocols discussed here are mainly the:

- Data-centric: based on queries, eliminate redundancy.
- Hierarchical: cluster the nodes, save energy.
- Location-based: use location to relay data to regions not to the whole network.

The aim of this paper is therefore describe the issues related to the architecture of the sensor networks system. It is also to identify the various implications of the process of data routing. Moreover, the data-centric routing will be discussed and focused in details.

## II. SYSTEM ARCHITECTURE AND DESIGN ISSUES

There are various system architectures as well as design aims and also limitations that have been set for the sensor networks. Following we have discussed the most critical sensor networks system architectures and design issues.

**2.1  Node Deployment:**

The node deployment can be deterministic. This means that the sensors are placed in a manual way and therefore the data is being routed via paths that are pre-determined. On the other hand, the node deployment can be self-organizing. This means that the nodes are rather dispersed at random like in ad hoc way (Sharef et al., 2013).

**2.2  Energy Considerations:**

The creation and setting up of routes while creating the infrastructure takes into consideration the energy issue. The radio transmits data and this is proportional to distance and also to the obstacles that exist in the route. For this reason establishing a multi-hop routing uses less energy compared to direct communication. If the nodes are close to the sink then direct communication is better. But since the nodes are dispersed here the multi-hop is the best option (Sahoo, Rath, and Puthal, 2012).

**2.3  Data Delivery Models:**

There are many forms for the data delivery model to the sink. They are:

- Continuous: data is sent periodically by the sensor
- Event-driven: data is transmitted only when an event happens
- Query-driven: data is transmitted only when a query takes place
- Hybrid: continuous + event-driven + query-driven delivery of the data.

The data delivery model influences the routing protocol. This can be clearly noticed when we think of energy consumption minimization and also the stability of the route (Prasan, and Murugappan, 2012).

**2.4  Node Capabilities:**

The sensor nodes have various functionalities within the sensor network. Previously, the sensor nodes were all considered rather homogeneous that have equal computational capacity, and also equal power and equal communication. A node, however, can be allotted just one function like sensing, relaying, or aggregation. This is because if the node does the three functions this might lead to the drainage of the node's energy. Some protocols rather use a cluster-head than a sensor because it is more powerful than the normal sensor in relation to energy, memory and also in relation to memory. The cluster-head is responsible for the aggregation and for the transmission to the sink (Blanchet, 2009).

When different than heterogeneous sensors are introduced this caused many data routing technical problems and issues. In this context, to monitor the temperature, the humidity, the pressure of the environment, as well as the detection of motion and image capture or video capture, many sensors have to be engaged. The sensors generate readings at various rates, and they face constraints related to quality, and they also have data delivery models that are different and multiple (Mondal and Sarddar, 2014).

**2.5  Data Aggregation/Fusion:**

As mentioned earlier, the nodes sometimes generate data that is rather redundant. For this reason, the similar packets transmitted from different nodes can be aggregated to reduce the number of transmissions. This is done through combining of the data using the suppression function (which means duplicates elimination). Other functions also can be used such as min, max, and average. Data computation uses less energy than transmission for this reason data aggregation is the ultimate tool. The data aggregation technique helps to save the energy efficiently as well as optimize the traffic (Casteran, and Filou, 2011) (Blanchet, 2009).

The aggregation functions are always allotted to nodes that are powerful and specialized in some of the networks. The aggregation of data is also possible through the use of certain techniques such as the signal processing technique. This is known as data fusion. Here the node can generate a signal that is much more accurate through the noise reduction and also via the use of beam-forming technique for the sake of combining the signals together (Sharef, Alaradi and Sharef, 2012) (Alipio and Tiglao 2017).

### III.    DATA-CENTRIC PROTOCOLS

It is sometimes not possible to designate and assign identifiers that are global for the nodes one by one. The deployment that is rather random of the sensor nodes renders it rather difficult to query given and specific nodes set. For this reason, the transmission of data is done from the sensor nodes inside each deployment area in a very big redundancy level. This is of course not efficient as it increases the energy consumption. Therefore, routing protocols have been selected for the sake of choosing certain sensor nodes and also for the sake of using data aggregation while relaying the data (Sharef, Alsaqour, and Ismail, 2013).

Data-centric routing therefore has been created. This is distinguished from the address-based routing that is rather traditional now. Queries in data-centric routing are sent to specific areas and the sensors in that specific area sends the data back to the sink. At this level, the first data-centric protocol used is SPIN. It checks the negotiation of the data between the nodes and it eliminates the redundancy and also it save the consumption of the energy. In later stages, the use of the Directed Diffusion was popular within the data-centric routing. Later, many other protocols were deployed. These protocols are discussed one by one in the following section (Aneja and Roy, 2016).

*3.1 Flooding and Gossiping*:

These mechanisms that are rather classical in approach and they are used to relay the data in sensor networks without any routing algorithms or topology maintenance. In flooding for example, the sensor that receives a data packet broadcasts it to all neighbors and this kind of process continues till the packet reaches either the destination or the maximum number of hops possible. Gossiping is another version of flooding but enhanced. Here the receiving node sends the packet to a neighbor that is randomly selected. This neighbor selects another neighbor, which picks another random neighbor to forward the packet to, and the process continues as such (Grumbach, and Wang, 2010).

It is important to mention that the implementation of flooding is easy. However, it has many disadvantages like:

- *Implosion:* duplicated messages sent to the same node cause impulsion.
- *Overlap:* this happens in case more than one node senses the same region and sends similar packets to the same neighbor
- *Resource blindness:* this means that the energy is being consumed with no attention paid to the energy constraints.

One positive side of gossiping is that Gossiping can simply eliminate the impulsion problem through the selection of a random node to send the packet rather than broadcasting. But in such a case there is always a delay in data propagation through the nodes. Figures 1 and 2 represent the impulsion and the overlap problems (Mondal and Sarddar, 2014).

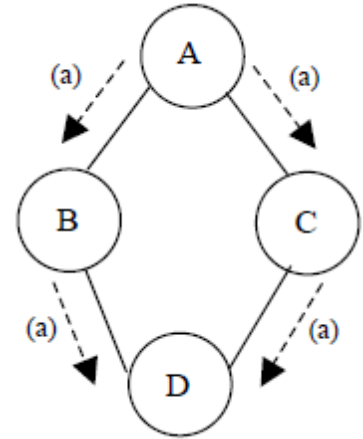

*Fig.1: The implosion problem. Noda A starts by flooding its data to all of its neighbors. D gets two same copies of data eventually, which is not necessary.*

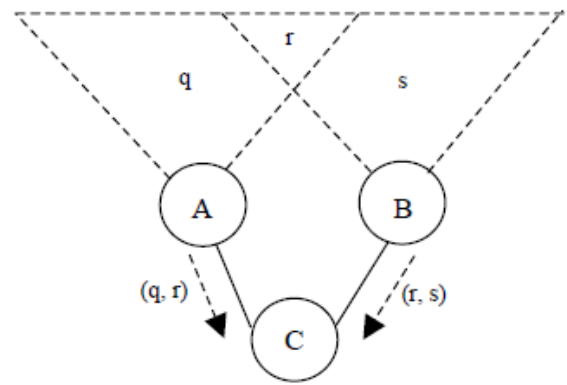

*Fig.2: The overlap problem. Two sensors cover an overlapping geographic region and C gets same copy of data from these sensors.*

*3.2 Sensor Protocols for Information via Negotiation (SPIN):*

The whole approach behind SPIN is giving a name to the data through the use of high level descriptors (meta-data). The exchange of meta-data between and among sensors through the data advertisement tool is done before the transmission. When the new data reaches the node it does an advertisement of that data to the neighbors. The neighbors that need the data can send a message to retrieve that data. The negotiation of the meta-data of SPIN is the solution for the problems of flooding like overlapping of sensing areas, the redundant passing of information, as well as the resource blindness. This way energy efficiency is achieved (Prasan, and Murugappan, 2012).

Meta-data has no standard format. It is rather related to specific applications. In SPIN, 3 messages are defined for exchanging the data in between the nodes:

- ADV: allows the sensor to advertise the meta-data that is particular
- REQ: that requests specific data.

- DATA: carries the data that is actual

The SPIN protocol is summarized in Figure 3.

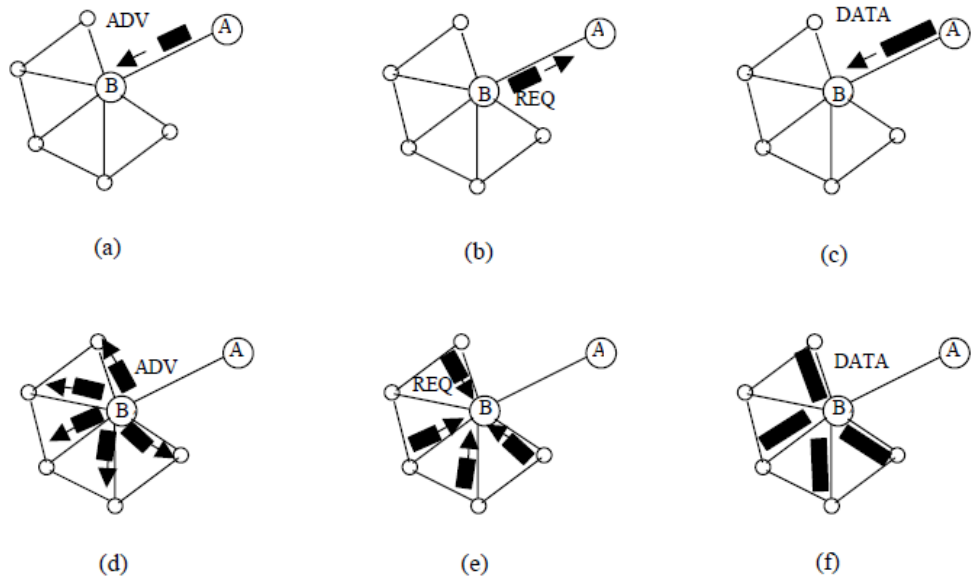

*Fig.3: SPIN Protocol. Node A starts by advertising its data to node B (a). Node B responds by sending a request to node A (b). After receiving the requested data (c), node B then sends out advertisements to its neighbors (d), who in turn send requests back to B (e-f).*

The SPIN protocol has many advantages. The changes that are topological are rather localized because the node does not need more than its neighbors that are single-hop. Moreover, SPIN is more efficient when it comes to energy saving (3.5 less than flooding). Also it reduces the data redundancy to half.

The problem with SPIN is that there is guarantee that the data will be delivered. For example, when the node that sends the request for certain needed data is located far away from the source node and also when the node in between is not interested in that particular data, then no data communication happens.

*3.3 Directed Diffusion:*

This is a very important aspect of the data-centric routing. The data is diffused via the sensor nodes through the use of a data name scheme. This is done to eliminate network layer operations that are not necessary to save the energy. If a node is interested in certain data, there must be a list of attribute-value pairs (interval, name of objects, duration, and area). Then the query is broadcast by the sink via the neighbors. The nodes that receive the interest cache it for using it later. Moreover, the nodes can establish a data aggregation that is qualified as in-network (Blanchet, 2009). The cached interests are used in the comparison of the data that is received against the values within the interest. There are also gradient fields within the interest itself. The gradient is simply the reply link to the neighbor that sent the interest. A gradient characterizes by the duration, data rate, and expiration time that have been retrieved from the fields of the interest. The use of interest and the gradients it is possible to create a path between the sources and the sink. Finally, the sink uses the path that is selected with smaller interval to resend the original interest message. The source node is therefore reinforced on that path to send the data frequently. Figure 4 summarizes the directed diffusion protocol phases (Grumbach, and Wang, 2010).

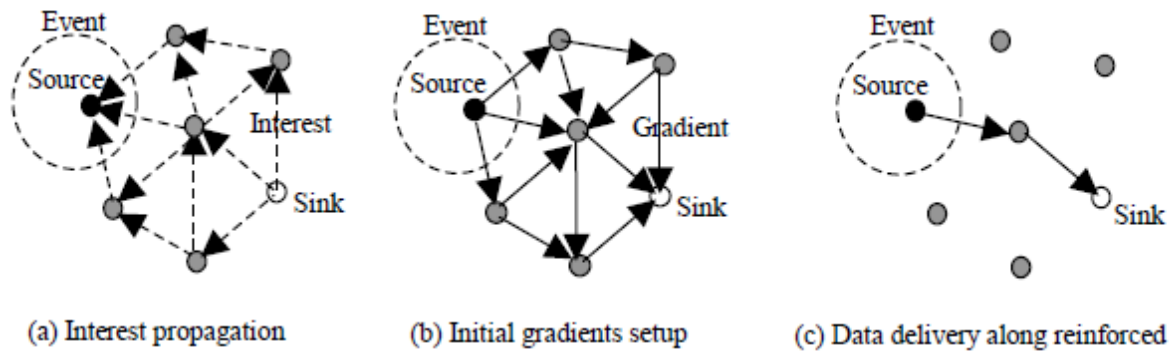

*Fig.4: Directed diffusion protocol phases*

Directed diffusion allows the path change and repair. If a path does not work another path is assigned. Therefore there should be a multitude of paths from the beginning assigned for the task. Directed Diffusion is different from SPIN as the sink flood tasks to the sensor nodes when some data is available, whereas SPIN makes the sensors advertise that data is available and the interested nodes can query the data. In directed diffusion contains communication that is neighbor-to-neighbor. There is no need for a mechanism that node addressing. The individual nodes can sense, cache and aggregate. Here caching helps the energy efficiency and also the delay. Direct Diffusion is energy efficient because of its on demand aspect. This means the global network topology maintaining is not needed. The problem with directed diffusion is that we cannot apply it to all the applications of the sensor network. This is because it has been based on a data delivery model that is query-driven (Mondal and Sarddar, 2014).

### 3.4 Energy-aware routing:
For the sake of prolonging the lifetime of network. Suggested that sub-optimal paths should be used. The paths are selected through the probability function. This depends on the amount of energy used by each path. The claim is that if we always use the minimum energy in the path this will cause the destruction of the nodes' energy using that path. One of the paths should be rather used with probability for the sake of increasing the lifetime of the whole network. This protocol has 3 phases:

a)  Setup phase: Localized flooding happens for the sake of finding routes and creating the routing tables.
b)  Data Communication Phase: Each node sends the packet via a random choice of a node from its

forwarding table through the use of the probabilities.
c)  Route maintenance phase: Localized flooding is not done frequently for the sake of keeping all paths alive.

Here one path is chosen at random among many alternatives for the sake of saving energy. Comparing to directed diffusion, this improves energy saving by 21.5%, and also increases network life by 44%. However, if a node or path fails there is no recovery, unlike the case of directed diffusion. Finally, compared to directed diffusion, gathering info about location and also setting nodes' addressing mechanism make the route setting complicated (Sahoo et al., 2012).

### 3.5 Rumor routing:
This is simply a variation of the directed diffusion. This works when there is no application for geographical routing criteria. When nodes observe an event the queries are routed to those nodes. There is no need to flood the whole network. In the event flooding through the network, the rumor routing algorithm uses agents (which are long-lived packets).

In case an event is detected by the node, the node adds the event to its table and also an agent is generated or created. The agents go through the network for the sake of spreading the information related to the events that are local to the far away nodes. In case an event query is generated by a node, the other nodes knowing the route reply through their event table referral. In this way there is no need to flood the whole network and in this way the cost is reduced. In comparison with directed diffusion, rumor routing uses one path only linking source to destination. This saves energy and also helps in nodes failure. The problem is that rumor routing only works well when the events are small in number.

*3.6  Gradient-Based Routing:*

There is another version of the directed diffusion. They call it the gradient-based routing (GBR). They suggest that keeping the number of hops in case of interest diffusion through the network is a good idea. The nodes therefore can have the height of the node, which is discovering a minimum number of hops to the sink. Here the gradient is the difference between the height of one node and the height of its neighbor node. The largest gradient is then sent on a link. In this context, there are 3 techniques for data spreading:

- Stochastic Scheme: in case of multiple next hops (having same gradient), node randomly chooses one.

- Energy-based scheme: in case the energy of a node goes decreases, its height goes up so other sensors do not send data to it.

- Stream-based scheme: new streams are diverted from nodes that are part of the other streams' path.

In this way, the data spreading scheme wants to establish an equal and even traffic distribution within the network as a whole. This aids load balancing on sensor nodes, and also helps in increasing the lifetime of the network.

*3.7  Constrained anisotropic diffusion routing (CADR):*

CADR is one of the protocols that is referred to as the general form of Directed Diffusion. Here, there are 2 main techniques (information-driven sensor querying (IDSQ)) and also (constrained anisotropic diffusion routing (CADR)). This is done for the sake of querying the sensors and routing the data within the network so as to increase the gain of info to a maximum, and at the same time decrease the delay and minimize the bandwidth. To achieve this, the sensors near a given event are activated and no others, and also the data routes are adjusted in a dynamic way. In the CADR case nodes do the info objective evaluation and the objective cost evaluation as well and then data is routed on the basis of the local information or cost gradient as well as the end-user requirement. For the case of IDSQ, the node that sends the query can decide which node is better in providing info needed and at the same time maintaining the energy cost balanced.  In this way, the CADR sends queries through selecting the sensor that should get the data, in this way it is better than directed diffusion in terms of energy saving (Shi et al., 2012).

*3.8  COUGAR*:

The COUGAR is basically a data-centric protocol considering the network like a distributed database system.

Here we utilize the queries that are declarative for the sake of abstracting the query processing from the network layer functions (like selecting the appropriate sensor). It is also for the sake of using data aggregation that is rather in-network in order to save the energy (Deng, Grumbach, and Monin, 2011).

The COUGAR protocol suggests one type of architecture to the sensor database system. Here the nodes choose a node as a leader that will aggregate and also transmit the data to the sink (gateway). Figure 5 describes the architecture. Here the sink generates a query plan. This query plan dictates the info related to the data flow and also about the in-network computation for the query that is incoming, and then forward that info to the appropriate and relevant nodes. This query plan gives a description of the way the leader is selected for the query (Qiu et al., 2017).

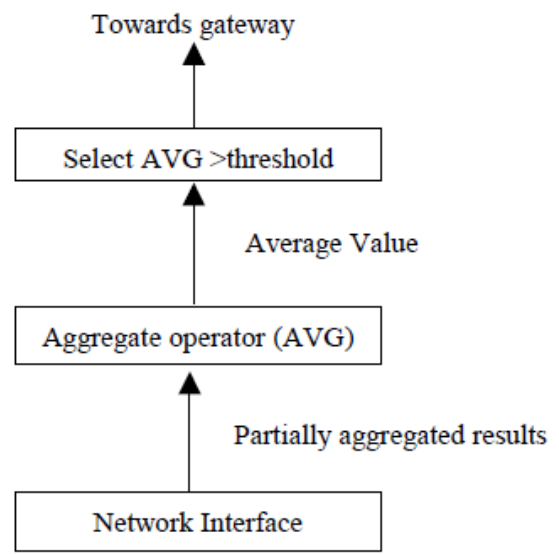

*Fig.5: Query plan at a leader one: The leader node gets all the readings, calculates the average and if it is greater than a threshold sends it to the gateway (sink).*

The COUGAR protocol gives an independent solution to the network layer to query the sensors. However, there a few disadvantages. When an additional query layer is added to the sensor node this causes additional overhead to sensor nodes related to storage as well as to energy consumption. When the data is computed from many nodes in an in-network manner, this needs synchronization, which means that the node that is relaying cannot send the data to the leader node until all packets are gathered from various sources (Shi et al., 2012).

*3.9  ACQUIRE*:

This is rather a mechanism for data-centric for the query of sensor networks that is rather new. This is called Active Query Forwarding in Sensor Networks (ACQUIRE). It regards the sensor network like a distributed database. It also considers it suitable for queries that rather complex (they are made out of many sub queries). Here the sink sends the query. The nodes that receive it use the pre-cached info to respond and to send the query to one other sensor. In case of out-dated cached info, the node assembles info from the other neighboring nodes. The query is sent back via the same path or through the shortest path to the sink when it is ready and resolved. This is a good approach as complex data queries get response from various nodes (Sahoo et al., 2012).

## IV.     CONCLUSION

The routing in sensor networks has been of great concern in comparison to the so-called traditional data routing in the rather wired networks. This paper has discussed and reviewed the research made on the data routing in sensor networks. It specifically discussed and described the data-centric category.

We call data-centric the protocols that name the data and also that query the nodes on the basis of data attributes. This is now followed by a lot of researchers as it enables the elimination of the overhead of forming clusters. It also eliminates the utilization of nodes that are rather specialized.

In the future, there will be the possibility of merging the sensor networks with the Internet (wired networks). The applications namely in security for example as well as for the monitoring of the environment need the transmission of the data that has been collected from the sensor nodes to the server so that it is further analyzed. One more aspect is that when the user makes a request to the sink this should be made through the use of the Internet.